# Adaptive Machine Learning for Time-Varying Systems: Towards 6D Phase Space Diagnostics of Short Intense Charged Particle Beams

Alexander Scheinker[1,a] and Spencer Gessner[2,b]

[1]Los Alamos National Laboratory, Los Alamos, NM 87545, USA
[2]SLAC National Accelerator Laboratory, Menlo Park, CA 94025, USA

March 22, 2022

**Abstract**
Particle accelerators are complex systems composed of coupled electromagnetic components including radio frequency resonant accelerating structures for acceleration and magnets for beam steering and transverse focusing. Charged particle beams are complex objects living in a six-dimensional phase space $(x,y,z,p_x,p_y,p_z)$. As bunches become shorter and more intense, the effects of nonlinear intra-bunch collective interactions such as space charge forces and bunch-to-bunch influences such as wakefields and coherent synchrotron radiation also increase. Shorter more intense bunches are also more difficult to accurately image because their dimensions are beyond the resolution of existing diagnostics and they may be destructive to intercepting diagnostics. The limited availability of detailed diagnostics for intense high energy beams is a fundamental challenge for the accelerator community because both beams and accelerators are time-varying systems that change in unpredictable ways. The detailed 6D distributions of beams emerging from sources vary with time due to factors such as evolving photocathode laser intensity profiles and the quantum efficiency of photocathodes. Accelerator magnets, RF amplifiers, and control systems are perturbed by external disturbances, beam-loading effects, temperature variations, and misalignments. Although machine learning (ML) methods have grown in popularity in the accelerator community in recently years, they are fundamentally limited when it comes to time-varying systems for which most current approaches are to simply collect large new data sets and perform re-training, something which is not feasible for busy accelerator user facilities because detailed beam measurements usually interrupt operations. New adaptive machine learning (AML) methods designed for time-varying systems are needed to aid in the diagnostics and control of high-intensity, ultrashort beams by combining deep learning tools such as convolutional neural network-based encoder-decoder architectures, model-independent feedback, physics constraints, and online models with real time non-invasive beam data, to provide a detailed virtual view of intense bunch dynamics.

[a]ascheink@lanl.gov, [b]sgess@slac.stanford.edu

**Background and Motivation**

A major challenge faced by advanced accelerators and in particular by those important to high energy physics (HEP), such as wakefield acceleration (WFA) experiments, is the ability to precisely generate and control the acceleration of extremely com- pressed (few fs/μm), high charge (few nC), high peak current (>200 kA) electron bunches with low energy-spread and tailored current profiles. In order to control the position-energy ($z$,$E$) 2D longitudinal phase space (LPS) of intense ultra-short bunches at high energy (>10 GeV) requires the ability to non-invasively measure their 2D LPS distributions at high resolution (< fs/μm). Furthermore, once they are generated, in order to precisely accelerate trains of closely spaced (ns) intense bunches requires the development of novel algorithms for sub-ns control of the electromagnetic fields of radio frequency (RF) accelerating cavities. Intense closely spaced bunches create strong wakefields in RF accelerating structures spoiling the emittance and acceleration of trailing bunches. This issue is particularly challenging for high-Q cryogenically cooled copper or superconducting accelerating structures, which are extremely efficient and narrow bandwidth. The precise control of ultra-short, intense, and closely spaced bunches in particle accelerators requires new adaptive machine learning (AML) algorithms for controls and non-invasive diagnostics. The 2019-20 HEP general accelerator R&D accelerator and beam physics (ABP) workshops identified four grand challenges which are the motivation for this work:

**1). Beam Intensity:** *How do we increase beam intensities by orders of magnitude?* This requires the development of low emittance sources of high intensity particle bunches, whose development would be greatly aided by more accurate diagnostics.

**2). Beam Prediction:** *How do we develop predictive "virtual particle accelerators"?* As bunch intensity increases, we need a combination of adaptive model-based non-invasive diagnostics coupled to real-time data from existing diagnostics to provide a virtual view of the 6D phase space of intense charged particle bunches. Such diagnostics can inform the design and development of higher intensity sources.

**3). Beam Control:** *How do we control the beam distribution down to the level of individual particles?* Utilizing a detailed virtual view of the beam, we can develop adaptive controls that automatically manipulate the 6D particle distribution. As beam control becomes more precise, it aids the development of new and more accurate virtual diagnostics by providing precise distributions at a certain accelerator location that can be used as input to a high-quality physics-based model for further predictions along the accelerator.

**4). Beam Quality:** *How do we increase beam phase-space density by orders of magnitude, towards quantum degeneracy limit?* Advanced diagnostics and controls will improve beam quality (preserving low emittance through higher levels of bunch compression) with active real-time adaptive feedback, and higher quality more predictable beams are in turn easier to control and to predict at higher energies downstream.

WFA techniques have the potential to accelerate beams within a few meters to the same energies that would require kilometers of traditional RF acceleration. WFAs can potentially serve as energy upgrades for an ILC, enable compact ILC designs, can be used to study extremely high-intensity and high-energy nonlinear beam dynamics, and have the potential to enable compact free electron lasers (FEL). For example, the Facility for Advanced Accelerator Experimental Tests (FACET) at SLAC has demonstrated acceleration of electrons[1] as well as positrons[2] to high energies within one meter of plasma. The AWAKE experiment at CERN uses transversely-focused (∼ 200μm), high-intensity ($3 \times 10^{11}$), high-energy (400 GeV) protons from CERN's Super Proton Synchrotron (SPS) accelerator to drive wakefields in a 10 meter-long plasma and accelerate electron bunches with MeV energy up to energies of 2 GeV[3]. FACET-II is currently being commissioned with the goal of providing custom tailored current profiles for various experiments with bunch lengths as low as (1 μm or ∼3 fs) and high peak currents (20 - 200 kA)[4].

The WFA process is extremely sensitive to the detailed longitudinal current profiles of these bunches and it would be of great benefit to have precise control over these profiles. However, the dynamics of extremely short and intense charged particle beams are difficult to control and quickly/accurately model due to collective effects such as space charge forces and wakefields. Furthermore, diagnostics are extremely limited for such high-current, high-energy, and short electron bunches. Even if lengthy, detailed measurements of beams are made and used as input into the models, due to uncertain and time-varying components and settings, the predictive power of the models drifts with time and quickly degrades. Therefore, currently WFA methods cannot produce beams that match the quality (such as emittance, energy spread, and reproducibility) of conventional accelerators.

[a]ascheink@lanl.gov, [b]sgess@slac.stanford.edu

**Proposed Innovation 1:** Adaptively-tuned Physics Models as Virtual Diagnostics: Virtual TCAV

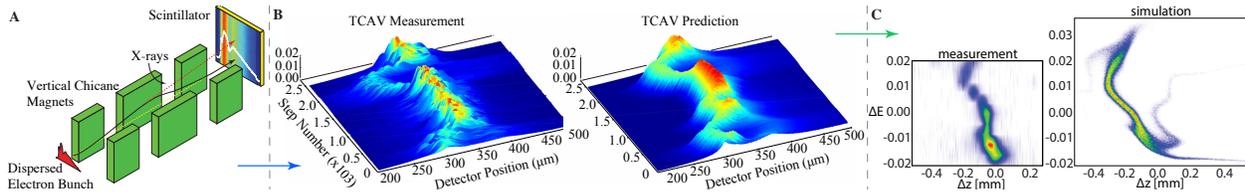

**Figure 1.** *Adaptive virtual diagnostic at FACET.* **A:** *Online model adaptively tuned to match energy spectrum prediction to non-invasive instrument.* **B:** *TCAV measurement predicted for time-varying beam.* **C:** *Virtual view of the longitudinal phase space[5].*

Machine learning (ML) methods can be used to learn complex relationships between coupled parameters and beam properties. Because both accelerator components and beams change with time, ML alone is not sufficient and requires the addition of adaptive feedback to automatically compensate for un-modeled disturbances and changes. We propose that adaptive ML techniques can be extremely useful for developing adaptive virtual diagnostics and adaptive feedback controls for shorter, more intense charged particle beams that are of importance to HEP science. One approach is to couple online physics-based models and data-based ML surrogate models together with model-independent adaptive feedback techniques from nonlinear feedback control theory that are by design robust to changes, nonlinearities, and external disturbances that cannot be accurately modeled[6–8]. The development of such a virtual diagnostic for WFA was first demonstrated at FACET[5], as shown in Figure 1, where an online model was adaptively tuned based on non-invasive diagnostics to provide a virtual TCAV measurement of the beam's longitudinal phase space (LPS). The ability of the virtual non-invasive diagnostic to track the beam's LPS was confirmed by simultaneously running a destructive TCAV measurement. Preliminary results towards developing an adaptive ML approach for LPS control were first demonstrated at the LCLS free electron laser where a neural network was trained to give instant estimates of parameter setting required for achieving a desired LPS distribution as measured by the TCAV following the undulator and then adaptive model-independent feedback was used to fine tune parameters and zoom in on and track the desired LPS distribution despite time-varying beam and accelerator parameters[9]. Adaptive model-independent feedback control was also demonstrated for real-time multi-objective optimization of CERN's AWAKE electron beam line for simultaneous orbit control and transverse emittance minimization[10].

[a]ascheink@lanl.gov, [b]sgess@slac.stanford.edu

**Proposed Innovation 2:** Adaptive ML for Time-Varying Systems: 6D Phase Space Diagnostics

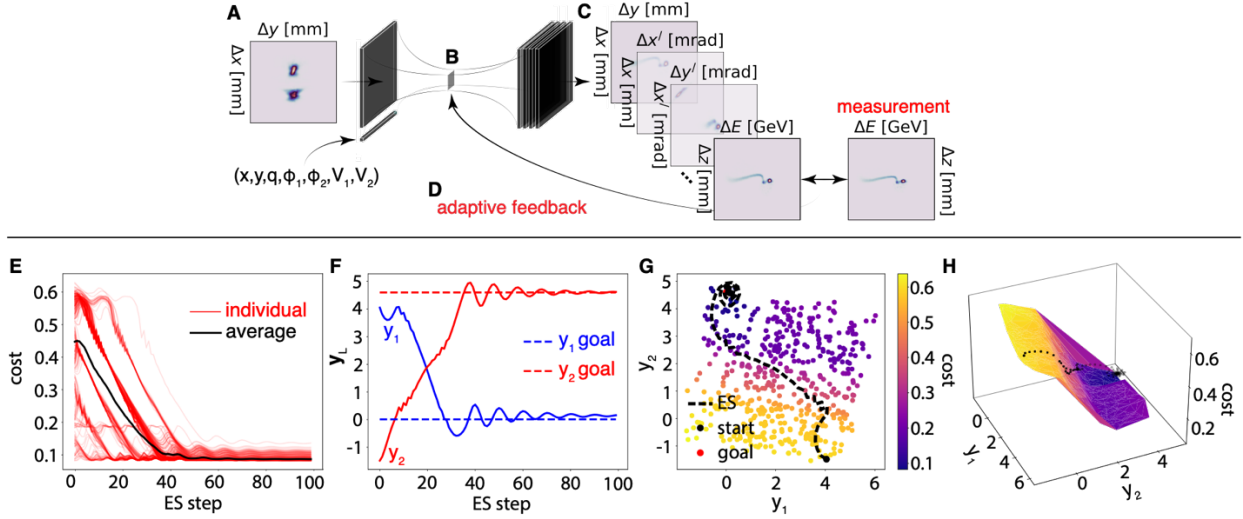

*Figure 2: Overview of an adaptive ML approach to virtual 6D diagnostics.* **A:** *An encoder-decoder convolutional neural network (CNN) takes high dimensional ($10^5$-$10^6$ dimensions) inputs of beam (x,y) distributions as well as scalar beam and accelerator quantities (input beam x and y offsets, charge, and phase and voltage of linac sections).* **B:** *The high dimensional inputs are mapped down to a low dimensional (2-10 dimensions) latent space representation.* **C:** *The generative half of the encoder-decoder then produces all 15 unique 2D projections of the beam's 6D phase space.* The overall encoder-decoder CNN going from steps (A)-(C) is trained via supervised learning.* **D:** *In application, the approach is un-supervised with input beam distributions and beam parameters assumed to be unknown and time-varying. Only a single 2D slice of the CNN's prediction, the longitudinal phase space, is compared to a TCAV-based measurement, and adaptive feedback is then used to adaptively tune the low-dimensional latent space.* **E:** *In this setup for 100 random initializations convergence was achieve on average within fewer than 50 iterations on the low-dimensional latent space.* **F:** *One example of convergence in the latent space is shown.* **G:** *One example of a path through the latent space is shown.* **H:** *The overall adaptive ML design allows us to create a convex cost function over the latent space to guarantee unique reconstructions. Figure adapted from[27].*

Various ML methods for particle accelerator applications are now being developed at facilities around the world. Methods have been developed to map accelerator parameter settings to LPS predictions[9,11], ML techniques have been developed to optimize FEL performance[12], a novel ghost imaging approach has been developed to map the time-varying quantum efficiency of photocathodes[13], multi-objective Gaussian process optimization has been applied to the nonlinear storage ring dynamics of SPEAR3[14], and various ML algorithms for identifying faulty beam position monitors and for optics corrections have been tested at CERN[15-17]. A broad overview of recent developments and the state of the art in adaptive controls and machine learning for particle accelerators can be found in the proceedings of the 2019 Advanced Control Methods for Particle Accelerators (ACM4PA) workshop[18]. In future work, the goal is to develop coupled adaptive ML-based controls and ML-based online diagnostics[19] that can utilize recently developed model-independent methods for the optimal control of unknown systems, such as accelerators and their beams[20], based on virtual diagnostics. The use of virtual diagnostics to guide automatic LPS control was recently studied in simulation for FACET-II[21]. Researchers at SLAC have developed diagnostics which utilize spectral measurements for increased resolution and prediction accuracy[22]. Researchers at DESY have been developing incredibly high resolution non-invasive longitudinal phase space (LPS) diagnostics utilizing convolutional neural networks[23]. At CERN ML tools have also been developed as virtual diagnostics for not just the accelerated beam, but for the accelerator itself, for example for identifying magnet errors based on beam measurements[24]. At CERN surrogate models have also been developed for fast simulation studies of the CLIC final focus system, mapping sextupole offsets to luminosity and beam sizes without requiring computationally expensive tracking and beam-beam simulations[25]. AT PSI researchers have been utilizing advanced polynomial chaos techniques to develop surrogate models which utilize methods for modeling stochastic differential equations with uncertainty quantification which can be used to construct global sensitivity models with error propagation and error analysis[26].

[a]ascheink@lanl.gov, [b]sgess@slac.stanford.edu

Despite recent advances in developing ML-based tools for particle accelerators, a major limitation is the application to time-varying systems, or systems with distribution shift, which are a major challenge for the machine learning community in general and are of particular importance to the particle accelerator community. If a system is changing with time, then a typical approach of learning a surrogate model requires continuous re-training to try and keep up with changing system characteristics. Although re-training is a feasible approach for certain problems, such as image recognition, in which acquiring new data does not interrupt operations, for particle accelerator ML applications invasive beam measurements can only be performed during dedicated development times so as not to interrupt operations. Furthermore even during dedicated experimental times detailed beam measurements such as quadrupole scan-based emittance measurements or wire scan-based profile measurements are very time consuming. At LANL efforts have been underway to develop adaptive ML tools which combine model-independent adaptive feedback techniques within the ML framework to make ML-based diagnostics and controls robust to uncertain time-variation of accelerators and their beams[27-29]. For example, as shown in Figure 2, a recent study has shown the potential to predict all 15 unique projections (x,y), … , (z,$p_z$), of a beam's 6D phase space by utilizing a convolutional neural network-based encoder-decoder generative network which is trained in a supervised learning approach and used in an adaptive un-supervised manner to track the time-varying 6D phase space of the beam without knowing the time-varying input beam distribution at the accelerator entrance.

Such adaptive ML methods have the potential to provide beyond state-of-the-art virtual diagnostics of the phase space of increasingly short and intense charged particle beams which will be very helpful in guiding feedback-based automatic tuning and control of the beam's phase space for detailed beam control such as fast and automatic tunning of custom current profiles over a wider range of beam properties (various beam energies, charges, bunch-to-bunch spacing, etc…) than has previously been demonstrated by initial adaptive ML methods such as those demonstrated for automated LPS tuning at the LCLS[9].

### References

[1] M. Litos, E. Adli, W. An, C. Clarke, C. Clayton, S. Corde, J. Delahaye, R. England, A. Fisher, J. Frederico, *et al.*, "High-efficiency acceleration of an electron beam in a plasma wakefield accelerator," Nature 515, 92 (2014). https://www.nature.com/articles/nature13882

[2] S. Corde, E. Adli, J. Allen, W. An, C. Clarke, C. Clayton, J. Delahaye, J. Frederico, S. Gessner, S. Green, *et al.*, "Multi-gigaelectronvolt acceleration of positrons in a self-loaded plasma wakefield," Nature 524, 442 (2015). https://www.nature.com/articles/nature14890

[3] E. Adli, A. Ahuja, O. Apsimon, R. Apsimon, A.-M. Bachmann, D. Barrientos, F. Batsch, J. Bauche, V. B. Olsen, M. Bernardini, *et al.*, "Acceleration of electrons in the plasma wakefield of a proton bunch," Nature 561, 363–367 (2018). https://www.nature.com/articles/s41586-018-0485-4

[4] C. Joshi, E. Adli, W. An, C. E. Clayton, S. Corde, S. Gessner, M. J. Hogan, M. Litos, W. Lu, K. A. Marsh, *et al.*, "Plasma wakefield acceleration experiments at FACET II," Plasma Physics and Controlled Fusion 60, 034001 (2018). https://iopscience.iop.org/article/10.1088/1361-6587/aaa2e3/meta

[5] A. Scheinker and S. Gessner, "Adaptive method for electron bunch profile prediction," Physical Review Special Topics-Accelerators and Beams 18, 102801 (2015), https://doi.org/10.1103/PhysRevSTAB.18.102801

[6] A. Scheinker, "Model independent beam tuning," in *Int. Partile Accelerator Conf.(IPAC'13), Shanghai, China, 19-24 May 2013* (JACOW Publishing, Geneva, Switzerland, 2013) pp. 1862–1864. http://accelconf.web.cern.ch/AccelConf/IPAC2013/papers/tupwa068.pdf?n=IPAC2013/papers/tupwa068.pdf

[7] A. Scheinker and D. Scheinker, "Bounded extremum seeking with discontinuous dithers," Automatica 69, 250–257 (2016). https://doi.org/10.1016/j.automatica.2016.02.023

[8] A. Scheinker and D. Scheinker, "Constrained extremum seeking stabilization of systems not affine in control," International Journal of Robust and Nonlinear Control 28, 568–581 (2018). https://doi.org/10.1002/rnc.3886

[9] A. Scheinker, A. Edelen, D. Bohler, C. Emma, and A. Lutman, "Demonstration of model-independent control of the longitudinal phase space of electron beams in the linac-coherent light source with femtosecond resolution," Physical review letters 121, 044801 (2018). https://doi.org/10.1103/PhysRevLett.121.044801

[a] ascheink@lanl.gov, [b] sgess@slac.stanford.edu

[10] A. Scheinker, S. Hirlaender, F. M. Velotti, S. Gessner, G. Z. Della Porta, V. Kain, B. Goddard, and R. Ramjiawan, "Online multi-objective particle accelerator optimization of the awake electron beam line for simultaneous emittance and orbit control," AIP Advances 10, 055320 (2020). https://doi.org/10.1063/5.0003423

[11] C. Emma, A. Edelen, M. Hogan, B. O'Shea, G. White, and V. Yakimenko, "Machine learning-based longitudinal phase space prediction of particle accelerators," Physical Review Accelerators and Beams 21, 112802 (2018). https://doi.org/10.1103/PhysRevAccelBeams.21.112802

[12] J. Duris, D. Kennedy, A. Hanuka, J. Shtalenkova, A. Edelen, P. Baxevanis, A. Egger, T. Cope, M. McIntire, S. Ermon, et al., "Bayesian optimization of a free-electron laser," Physical Review Letters 124, 124801 (2020). https://doi.org/10.1103/PhysRevLett.124.124801

[13] K. Kabra, S. Li, F. Cropp, T. J. Lane, P. Musumeci, and D. Ratner, "Mapping photocathode quantum efficiency with ghost imaging," Physical Review Accelerators and Beams 23, 022803 (2020). https://doi.org/10.1103/PhysRevAccelBeams.23.022803

[14] M. Song, X. Huang, L. Spentzouris, and Z. Zhang, "Storage ring nonlinear dynamics optimization with multi-objective multi-generation gaussian process optimizer," Nuclear Instruments and Methods in Physics Research Section A: Accelerators, Spectrometers, Detectors and Associated Equipment , 164273 (2020). https://doi.org/10.1016/j.nima.2020.164273

[15] E. Fol, Evaluation of machine learning methods for LHC optics measurements and corrections software, Ph.D. thesis, Hochschule, Eng. Econ., Karlsruhe (2017). http://cds.cern.ch/record/2309558/files/CERN-THESIS-2017-336.pdf?version=1

[16] E. Fol, J. M. Coello de Portugal, and R. Tomás, "Jacow: Application of machine learning to beam diagnostics," (2019). http://cds.cern.ch/record/2716703/files/tuoa02.pdf?version=1

[17] E. Fol, J. C. de Portugal, G. Franchetti, and R. Tomás, "Optics corrections using machine learning in the lhc," in Proceedings of the 2019 International Particle Accelerator Conference, Melbourne, Australia (2019). https://accelconf.web.cern.ch/ipac2019/papers/thprb077.pdf

[18] A. Scheinker, C. Emma, A. Edelen, and S. Gessner, "Advanced control methods for particle accelerators (acm4pa) 2019," Tech. Rep. (Los Alamos National Lab. (LANL), Los Alamos, NM (United States), 2019. https://permalink.lanl.gov/object/tr?what=info:lanl-repo/lareport/LA-UR-19-32526

[19] A. Edelen, N. Neveu, M. Frey, Y. Huber, C. Mayes, and A. Adelmann, "Machine learning for orders of magnitude speedup in multiobjective optimization of particle accelerator systems," Physical Review Accelerators and Beams 23, 044601 (2020). https://doi.org/10.1103/PhysRevAccelBeams.23.044601

[20] A. Scheinker and D. Scheinker, "Extremum seeking for optimal control problems with unknown time-varying systems and unknown objective functions," International Journal of Adaptive Control and Signal Processing (2020). https://doi.org/10.1002/acs.3097

[21] A. Scheinker, S. Gessner, C. Emma, and A. L. Edelen, "Adaptive model tuning studies for non-invasive diagnostics and feedback control of plasma wakefield acceleration at FACET-II," Nuclear Instruments and Methods in Physics Research Section A: Accelerators, Spectrometers, Detectors and Associated Equipment , 163902 (2020). https://doi.org/10.1016/j.nima.2020.163902

[22] A. Hanuka, et al. "Accurate and confident prediction of electron beam longitudinal properties using spectral virtual diagnostics." Scientific Reports 11.1 (2021): 1-10. https://doi.org/10.1038/s41598-021-82473-0

[23] J. Zhu, et al. "High-fidelity prediction of megapixel longitudinal phase-space images of electron beams using encoder-decoder neural networks." Physical Review Applied 16.2 (2021): 024005.

[24] E. Fol, et al. "Supervised learning-based reconstruction of magnet errors in circular accelerators." The European Physical Journal Plus 136.4 (2021): 365. https://doi.org/10.1140/epjp/s13360-021-01348-5

[25] J. Ögren, C. Gohil, and D. Schulte. "Surrogate modeling of the CLIC final-focus system using artificial neural networks." Journal of Instrumentation 16.05 (2021): P05012. https://doi.org/10.1088/1748-0221/16/05/P05012

[26] A. Adelmann. "On nonintrusive uncertainty quantification and surrogate model construction in particle accelerator modeling." SIAM/ASA Journal on Uncertainty Quantification 7.2 (2019): 383-416. https://doi.org/10.1137/16M1061928

[27] A. Scheinker. "Adaptive machine learning for time-varying systems: low dimensional latent space tuning." Journal of Instrumentation 16.10 (2021): P10008. https://doi.org/10.1088/1748-0221/16/10/P10008

[28] A. Scheinker, et al. "An adaptive approach to machine learning for compact particle accelerators." Scientific reports, 11(1), 1-11 (2021). https://doi.org/10.1038/s41598-021-98785-0

[29] A. Scheinker. "Adaptive Machine Learning for Robust Diagnostics and Control of Time-Varying Particle Accelerator Components and Beams." Information 12.4 (2021): 161. https://doi.org/10.3390/info12040161

[a]ascheink@lanl.gov, [b]sgess@slac.stanford.edu